\documentclass[prd,nofootinbib,reprint,twocolumn,showpacs,superscriptaddress,floatfix]{revtex4-1}
\pdfoutput=1 
\usepackage{bm, color} 
\usepackage{amssymb,amsfonts,slashed,amsthm,amsmath, graphicx, soul, empheq, cancel, xcolor}
\usepackage{hyperref}
\usepackage[caption=false]{subfig}
\begin{document}

\newcommand{\vev}[1]{ \left\langle {#1} \right\rangle }
\newcommand{\bra}[1]{ \langle {#1} | }
\newcommand{\ket}[1]{ | {#1} \rangle }
\newcommand{\eV}{ \ {\rm eV} }
\newcommand{\KeV}{ \ {\rm keV} }
\newcommand{\MeV}{\  {\rm MeV} }
\newcommand{\GeV}{\  {\rm GeV} }
\newcommand{\TeV}{\  {\rm TeV} }
\newcommand{\1}{\mbox{1}\hspace{-0.25em}\mbox{l}}
\newcommand{\Red}[1]{{\color{red} {#1}}}

\newcommand{\lmk}{\left(}  
\newcommand{\rmk}{\right)}
\newcommand{\lkk}{\left[}  
\newcommand{\rkk}{\right]}
\newcommand{\lhk}{\left \{ }  
\newcommand{\rhk}{\right \} }
\newcommand{\del}{\partial}  
\newcommand{\la}{\left\langle} 
\newcommand{\ra}{\right\rangle}
\newcommand{\half}{\frac{1}{2}}

\newcommand{\bea}{\begin{array}}
\newcommand{\eea}{\end{array}}
\newcommand{\beq}{\begin{eqnarray}}
\newcommand{\eeq}{\end{eqnarray}}
\newcommand{\eq}[1]{Eq.~(\ref{#1})}

\newcommand{\dd}{\mathrm{d}}
\newcommand{\Mpl}{M_{\rm Pl}}
\newcommand{\mg}{m_{3/2}}
\newcommand{\abs}[1]{\left\vert {#1} \right\vert}
\newcommand{\mphi}{m_{\phi}}
\newcommand{\Hz}{\ {\rm Hz}}
\newcommand{\for}{\quad \text{for }}
\newcommand{\Min}{\text{Min}}
\newcommand{\Max}{\text{Max}}
\newcommand{\Kahler}{K\"{a}hler }
\newcommand{\cphi}{\varphi}
\newcommand{\Tr}{\text{Tr}}
\newcommand{\diag}{{\rm diag}}

\newcommand{\SUf}{SU(3)_{\rm f}}
\newcommand{\Upq}{U(1)_{\rm PQ}}
\newcommand{\Zpq}{Z^{\rm PQ}_3}
\newcommand{\Cpq}{C_{\rm PQ}}
\newcommand{\ubar}{u^c}
\newcommand{\dbar}{d^c}
\newcommand{\ebar}{e^c}
\newcommand{\nubar}{\nu^c}
\newcommand{\Ndw}{N_{\rm DW}}
\newcommand{\Fpq}{F_{\rm PQ}}
\newcommand{\fpq}{v_{\rm PQ}}
\newcommand{\Br}{{\rm Br}}
\newcommand{\Lag}{\mathcal{L}}
\newcommand{\Lqcd}{\Lambda_{\rm QCD}}

\newcommand{\ji}{j_{\rm inf}} 
\newcommand{\jb}{j_{B-L}} 
\newcommand{\M}{M} 
\newcommand{\im}{{\rm Im} }
\newcommand{\re}{{\rm Re} }

\def\lrf#1#2{ \left(\frac{#1}{#2}\right)}
\def\lrfp#1#2#3{ \left(\frac{#1}{#2} \right)^{#3}}
\def\lrp#1#2{\left( #1 \right)^{#2}}
\def\REF#1{Ref.~\cite{#1}}
\def\SEC#1{Sec.~\ref{#1}}
\def\FIG#1{Fig.~\ref{#1}}
\def\EQ#1{Eq.~(\ref{#1})}
\def\EQS#1{Eqs.~(\ref{#1})}
\def\TEV#1{10^{#1}{\rm\,TeV}}
\def\GEV#1{10^{#1}{\rm\,GeV}}
\def\MEV#1{10^{#1}{\rm\,MeV}}
\def\KEV#1{10^{#1}{\rm\,keV}}
\def\blue#1{\textcolor{blue}{#1}}
\def\red#1{\textcolor{blue}{#1}}

\newcommand{\eff}{\Delta N_{\rm eff}}
\newcommand{\neff}{\Delta N_{\rm eff}}
\newcommand{\cc}{\Omega_\Lambda}
\newcommand{\Mpc}{\ {\rm Mpc}}
\newcommand{\Msolar}{M_\odot}

\def\sn#1{\textcolor{red}{#1}}
\def\SN#1{\textcolor{red}{[{\bf SN:} #1]}}
\def\SYH#1{\textcolor{blue}{[{\bf SYH:} #1]}}

\definecolor{purple}{HTML}{A020F0}
\def\FL#1{\textcolor{purple}{[{\bf FL:} #1]}}

\title{Transient Bias for CP Domain Wall Decay and Dark Matter}

\author{Sally Yuxuan Hao}
\affiliation{Tsung-Dao Lee Institute, Shanghai Jiao Tong University, \\
No.~1 Lisuo Road, Pudong New Area, Shanghai 201210, China}
\affiliation{School of Physics and Astronomy, Shanghai Jiao Tong University, \\
800 Dongchuan Road, Shanghai 200240, China}

\author{Fangchao Liu}
\affiliation{Tsung-Dao Lee Institute, Shanghai Jiao Tong University, \\
No.~1 Lisuo Road, Pudong New Area, Shanghai 201210, China}
\affiliation{School of Physics and Astronomy, Shanghai Jiao Tong University, \\
800 Dongchuan Road, Shanghai 200240, China}

\author{Shota Nakagawa}
\affiliation{Tsung-Dao Lee Institute, Shanghai Jiao Tong University, \\
No.~1 Lisuo Road, Pudong New Area, Shanghai 201210, China}
\affiliation{School of Physics and Astronomy, Shanghai Jiao Tong University, \\
800 Dongchuan Road, Shanghai 200240, China}

\author{Yuichiro Nakai}
\affiliation{Tsung-Dao Lee Institute, Shanghai Jiao Tong University, \\
No.~1 Lisuo Road, Pudong New Area, Shanghai 201210, China}
\affiliation{School of Physics and Astronomy, Shanghai Jiao Tong University, \\
800 Dongchuan Road, Shanghai 200240, China}

\begin{abstract}

Spontaneous CP violation (SCPV) provides an attractive solution to the strong CP problem. However, SCPV after inflation suffers from the formation of CP domain walls, requiring the maximal temperature of the Universe to lie below the CP-breaking scale. In the present work, we then propose a dynamical mechanism that removes this cosmological constraint without introducing permanent explicit CPV. We consider a new scalar field that acquires a large field value with a nontrivial phase in the early Universe and induces a transient bias among degenerate CP vacua through a higher-dimensional interaction with a CP-breaking scalar field. This bias triggers the decay of CP domain walls after they form. As the new scalar field evolves toward the origin, the bias disappears, leaving the low-energy CP structure intact. We derive the conditions for successful domain wall decay and identify the viable parameter space. Furthermore, we point out that the coherent oscillation of the new scalar field naturally survives as dark matter, linking the resolution of the CP domain wall problem to the origin of dark matter.

\end{abstract}

\maketitle

\section{Introduction
\label{sec:introduction}}

The absence of observable CP violation in the strong interaction remains one of the most intriguing puzzles in modern particle physics. In the Standard Model (SM), the Lagrangian of QCD contains the CP-violating $\bar \theta$ parameter,
which is constrained to satisfy $|\bar\theta| \lesssim 10^{-10}$ by the neutron electric dipole moment (EDM) measurement
\cite{Baker:2006ts,Pendlebury:2015lrz}. Explaining the extreme smallness of $\bar\theta$ constitutes the strong CP problem.

A promising approach to the problem is to impose CP as a fundamental symmetry of the underlying theory and break it spontaneously. In such a framework of spontaneous CP violation (SCPV), $\bar\theta$ vanishes at the fundamental level, while the observed CP violation in the Cabibbo-Kobayashi-Maskawa (CKM) matrix arises from complex vacuum expectation values (VEVs) of scalar fields
\cite{Nelson:1983zb,Barr:1984qx,Barr:1984fh,Bento:1991ez,Barr:1993hb,Dine:1993qm,Hiller:2001qg,Hiller:2002um}
(for more recent works, see e.g. Refs.~\cite{Vecchi:2014hpa,Dine:2015jga,Davidi:2017gir,Evans:2020vil,Valenti:2021xjp,Fujikura:2022sot,Girmohanta:2022giy,Asadi:2022vys,Bai:2022nat,Feruglio:2023uof,Dine:2024bxv,Nakagawa:2024ddd,Feruglio:2024ytl,Murai:2024alz,Ferro-Hernandez:2024snl,Jiang:2024frx,Feruglio:2024dnc,Murai:2024bjy,Feruglio:2025ajb,Liu:2025ycm}).

Despite the theoretical attractiveness, SCPV faces an important cosmological challenge. If SCPV occurs after inflation, different Hubble patches can settle into different CP vacua, which are degenerate in energy and related by the CP transformation, leading to the formation of CP domain walls. The problem is that such domain walls can be stable and eventually dominate the energy density of the Universe.
Consequently, it has been usually assumed that CP symmetry is broken before or during inflation and never restored thereafter. This requirement imposes an upper bound on the maximal temperature after inflation, and significantly restricts the thermal history of the Universe,
in particular scenarios of baryogenesis such as the thermal leptogenesis
\cite{Fukugita:1986hr}.

A straightforward way to eliminate CP domain walls is to introduce an explicit CP-breaking effect that lifts the vacuum degeneracy. However, this approach generically threatens the solution to the strong CP problem, as the explicit CPV can induce a large contribution to $\bar\theta$. Moreover, if CP is a gauge symmetry
\cite{Dine:1992ya,Choi:1992xp}, explicit CP-breaking operators are not expected to be present. It is therefore desirable to find a mechanism that removes CP domain walls while preserving the CP structure responsible for solving the strong CP problem.

In the present paper, we investigate a dynamical mechanism to remove CP domain walls without introducing a permanent
explicit CPV. We introduce a new scalar field that acquires a large field value with a nontrivial phase in the early Universe. Through a higher-dimensional interaction with the CP-breaking field, the new scalar field induces a transient bias among the degenerate CP vacua. This bias can generate a pressure difference across CP domain walls and then trigger their decay. As the Universe expands, the new scalar field evolves toward the origin so that the induced bias disappears. The mechanism therefore dynamically mimics an explicit CPV during a limited cosmological epoch without leaving a permanent imprint on the underlying theory.\footnote{The similar dynamical mechanism has been explored for the axion domain wall decay
\cite{Ibe:2019yew,Hao:2025kcz}.}

An additional feature of the scenario is that the new scalar field naturally survives as dark matter (DM). As the temperature decreases, the scalar field begins coherent oscillation around its minimum, while the transient bias becomes ineffective. We will show that the resulting relic abundance can account for the observed DM density in a viable region of the parameter space. In this way, the same field is responsible for both the elimination of CP domain walls and the origin of DM.

The rest of the paper is organized as follows. In Sec.~\ref{sec:light}, we introduce our setup with a new scalar field coupled to a CP-breaking field through a higher-dimensional interaction and explore the cosmological evolution of their radial modes.
The evolution of the phase modes are discussed in Sec.~\ref{subsec:eta}.
Then, Sec.~\ref{domain_wall} clarifies conditions for successful CP domain wall decay. In Sec.~\ref{sec_DM}, we investigate the abundance of DM originated from the coherent oscillation of the new scalar field and identify a viable model parameter space. In Sec.~\ref{sec_UV}, a possible UV completion of the setup is described. Sec.~\ref{sec:Discussion} is devoted to conclusions and discussions.

\section{Radial mode dynamics
\label{sec:light}}

The key ingredient of our mechanism is a light scalar field $S$ that carries a large homogeneous complex field value during inflation. When a CP-breaking scalar field $\eta$ later acquires a complex VEV and forms CP domain walls, a mixing interaction with $S$ generates a temporary bias term that lifts the vacuum degeneracy and drives the domain walls to collapse. Subsequently, as the Hubble parameter becomes smaller than the mass of $S$, $H \lesssim m_S$, $S$ begins oscillating around the origin and the induced CP violation disappears. In this way, the CP domain wall problem is solved dynamically without requiring permanent explicit CP breaking.
We start with the description of our setup and study the cosmological evolution of the two complex scalar fields, $S$ and $\eta$. In particular, we here focus on their radial modes.

\subsection{Setup}

Let us introduce a light complex scalar field $S$, which has a higher-dimensional interaction with a CP-breaking scalar field $\eta$.
We assume that these complex scalar fields have the following CP-symmetric potential:
\beq
\begin{aligned}
V(\eta,S) &\supset V(\eta) + \frac{1}{(n!)^2} \frac{\lambda_S^2}{M_{\mathrm{Pl}}^{2n-4}}|S|^{2n} + m_S^2|S|^2\\
&\quad + \lmk \frac{\lambda}{m! \, \ell! \, M_{\mathrm{Pl}}^{m+\ell-4}} S^m \eta^\ell + \mathrm{h.c.}\rmk ,
\end{aligned}
\label{Vtot}
\eeq
where $V(\eta)$ is given by
\beq
V(\eta) =\lambda_\eta\lmk|\eta|^2-\frac{v_{\rm CP}^2}{2}\rmk^2 + (\alpha \eta^2 +\beta \eta^4+ \text{h.c.}) \ .
\label{Veta}
\eeq
Here, $\lambda_\eta,\lambda_S,\lambda, \beta$ denote real dimensionless coupling constants, $m_S^2$ and $\alpha$ are real mass-squared parameters, $\Mpl\equiv1/\sqrt{8\pi G}$ with the Newton constant $G$ is the reduced Planck scale, and $v_{\rm CP}$ represents the scale of spontaneous CP symmetry breaking. 
The theory has a flat direction, $|\eta|\sim v_{\rm CP}$, which is only deformed by the second and third term.

We impose a $Z_2$ symmetry, under which both $\eta$ and $S$ are odd. 
This requires $\ell + m$ to be even in Eq.~\eqref{Vtot}.
We will address the rather specific form of the potential \eqref{Vtot} by considering a possible UV completion in Sec.~\ref{sec_UV}.
Note that Eq.~\eqref{Veta} is also not the most general form for $\eta$, as terms like $|\eta|^2\eta^2+ \rm h.c.$ are  possible. However, since such terms do not qualitatively change our discussion,
we do not include them for simplicity.

\subsection{Evolution of the $|S|$ field
\label{sec:evolve_S}}

\subsubsection{Inflation era}

The essence of our mechanism for CP domain wall decay lies in the assumption that the scalar $S$ acquires a large field value during inflation. In this epoch, the total energy density is dominated by the inflaton potential, $V_{\rm inf}\simeq 3H_{\rm inf}^2 M_{\rm Pl}^2$ ($H_{\rm inf}$ is the Hubble parameter during inflation), which induces a negative mass term for $S$. The relevant scalar potential involving the inflaton $\phi$ takes the form \cite{Harigaya:2015hha,Ibe:2019yew,Hao:2025kcz},
\begin{align}
V(\phi,\eta,S) \supset V_{\rm inf}(\phi) 
+ \frac{c_\eta}{3}\frac{V_{\rm inf}(\phi)}{M_{\rm Pl}^2}|\eta|^2 
- \frac{c_S}{3}\frac{V_{\rm inf}(\phi)}{M_{\rm Pl}^2}|S|^2 ,
\label{Vinf}
\end{align}
where $c_\eta$ and $c_S$ represent positive constants.
We assume that the inflaton $\phi$ decays predominantly into the SM particles, and the scalar $S$ couples to the rest of the theory only through Planck suppressed terms, so that $S$ is not thermalized. 
For $c_S H_{\rm inf}^2 \gg m_S^2$, the potential for $S$ in this era is approximately given by
\beq
V(S) \simeq \frac{1}{(n!)^2} \frac{\lambda_S^2}{M_{\rm Pl}^{2n-4}}|S|^{2n} - c_S H_{\rm inf}^2 |S|^2,
\eeq
which determines the expectation value of the radial component of $S$ during inflation:
\beq
\langle |S_{\rm inf}| \rangle \simeq \left(\sqrt{\frac{c_S}{n}}\frac{n!}{\lambda_S}\right)^{\frac{1}{n-1}} \left(\frac{H_{\rm inf}}{M_{\rm Pl}}\right)^{\frac{1}{n-1}} M_{\rm Pl} \ .
\label{Sinf}
\eeq
For instance, taking $H_{\rm inf}=10^{12}\GeV$, $n=6$, and $c_S=1$, one needs $\lambda_S\gtrsim10^{-4}$ for a sub-Planckian VEV of $S$, $\langle |S_{\rm inf}|\rangle \lesssim \Mpl$.
In general, $\langle S_{\rm inf} \rangle$ carries a physical initial complex phase $\theta_S$.

\subsubsection{Radiation-domination era}

The radiation-domination (RD) era follows the inflation era.
In the present paper, we simply assume the instantaneous reheating process,\footnote{The intermediate epoch can be dominated by the inflaton oscillation, in which $H\propto T^{3/2}$.
We can also utilize the scaling solution of $S$ in this case \cite{Ibe:2019yew,Hao:2025kcz}.}
where the temperature after the completion of reheating is given by 
\begin{align}
T_\text{RH} &\sim \lmk\frac{90}{\pi^2g_*}\rmk^{\frac{1}{4}} \sqrt{M_\text{Pl}H_\text{inf}}\nonumber\\
&\simeq 10^{15} \GeV \lmk\frac{H_{\rm inf}}{10^{12}\GeV}\rmk^{\frac{1}{2}}.
\end{align}
We assume that $m_S$ is much smaller than $H$ at the beginning of the RD era. The equation of motion (EOM) of $|S|$ is then given by
\beq
\ddot{|S|} + 3H \dot{|S|} + \frac{n\lambda_S^2}{2^{n-1} (n!)^2 M_{\text{Pl}}^{2n-4}} |S|^{2n-1} = 0 \ .
\eeq
We then obtain a value of $|S|$ during the RD era for $n \geq 6$ by using the scaling solution \cite{Harigaya:2015hha,Ibe:2019yew,Hao:2025kcz},
\begin{align} \label{SRD}
\langle |S_{\text{RD}}| \rangle \simeq \left[ \frac{2(n-3)(n!)^2}{n(n-1)^2 \lambda_{{S}}^2}  \right]^{\frac{1}{2(n-1)}} \left( \frac{H}{M_{\text{Pl}}}\right)^{\frac{1}{n-1} }M_{\text{Pl}} \ .
\end{align}
Note that the case of $n = 5$ also leads to a large value of $S$, but its behavior is different and more complicated. Thus, we choose $n = 6$ as a reference value throughout the present paper.

When $m_S \sim H$, the $S$ field starts to oscillate around the origin, so that the induced CP violation disappears.
Since this CP violation effect should be maintained until the collapse of CP domain walls or $T\sim v_{\rm CP}$,
we find an upper bound on $m_S$,
\beq
\begin{aligned}
m_S &\lesssim \sqrt{\frac{\pi^2g_*}{90}}\frac{v^2_{\rm CP}}{\Mpl}\\
& \simeq 1\GeV \lmk\frac{g_*}{106.75}\rmk^{\frac{1}{2}}\lmk\frac{v_{\rm CP}}{10^{9}\GeV}\rmk^2 , 
\end{aligned}
\eeq
where $g_*$ denotes the effective number of relativistic degrees of freedom.

\subsection{Evolution of the $|\eta|$ field
\label{eta_evolve}}

During inflation, the potential of $\eta$ is given by
\begin{align}
V_\eta^{\text{inf}} = \, &(c_\eta H_{\text{inf}}^2-\lambda_\eta v_{\text{CP}}^2)|\eta|^2 + \lambda_\eta |\eta|^4 + \alpha \eta^2 + \beta \eta^4 \notag \\
&+ \frac{\lambda \langle S_\text{inf} \rangle^m }{m!\ell! M_{\text{Pl}}^{m+\ell-4}} \eta^\ell + \text{h.c.}+\text{const.}, \end{align}
where we assume that $H_{\text{inf}} \gg v_{\text{CP}}$, so that the quadratic term is positive.
As $\lambda_\eta>0$, the vacuum structure depends on the mixing term, in other words, on the value of $\ell$.
For $\ell=1$, a single minimum deviates from the origin due to the linear term. 
For $\ell=2,3$, and $4$, there are either one or two local minima along the radial direction, and depending on the parameters, the global minimum can either sit at zero or a nonzero value.
For $\ell = 5,7,9, ...$,  the potential is unbounded from below, unless an extra higher-dimensional $(>\ell)$ term is introduced. 
Therefore, we consider $\ell = 1, 2, 3, 4$.

Let us focus on the case of $\ell=1$, which has a single minimum for sure. The balancing between the effective linear term from $\lambda\langle S_\text{inf} \rangle^m \eta$ and the quartic term $\lambda_\eta|\eta|^4$ determines the VEV,
\begin{equation}
    \langle|\eta_\text{inf}|\rangle \simeq \left(\frac{\lambda}{4\lambda_\eta m!} \right)^{\frac{1}{3}}\left(\frac{\langle S_\text{inf}\rangle}{M_\text{Pl}}\right)^{\frac{m}{3}}M_\text{Pl} \ .
\end{equation}
As we will see in the following, the minimal model which successfully solves the domain wall problem is the one with $m = 11$ and $n =6$.
In this case, $\langle |\eta_\text{inf}|\rangle$ is rather large during inflation,
\beq \label{etainf}
    \langle |\eta_\text{inf}| \rangle &\simeq& 4.4 \times 10^{14} \GeV \lmk\frac{\lambda_\eta/\lambda}{10}\rmk^{-\frac{1}{3}}\nonumber\\
    &\times&\lmk\frac{\lambda_S}{10^{-3}}\rmk^{-\frac{11}{15}}\lmk\frac{H_{\rm inf}}{10^{12}\GeV}\rmk^{\frac{11}{15}}.
\eeq

After the Universe reheats at $T > v_\text{CP}$, $\eta$ acquires the thermal mass of $\mathcal{O}(T^2)$ through the coupling with the SM sector, and $|\eta| $ is settled to
\begin{align} \label{thermalETAVEV}
     \langle|\eta|\rangle &\simeq \frac{\lambda}{m!T^2M_\text{Pl}^{m-3}} \langle |S_\text{RD}|\rangle^m  \notag \\
     &\simeq \frac{\lambda }{m!} \left[ \frac{\pi^2g_*}{90} \frac{2(n-3)(n!)^2}{n(n-1)^2\lambda_S^2} \right]^{\frac{m}{2(n-1)}} \left(\frac{T}{M_\text{Pl}} \right)^{\frac{2m}{n-1}-2} M_\text{Pl} .
\end{align}
As the Universe cools down, this expectation value gets much smaller than $v_\text{CP}$ at $T\sim v_{\rm CP}$. 
Thus, we can consider that $\eta$ is stabilized at the origin by the thermal mass term at around that temperature.\footnote{Even for a high reheating temperature, non-restored CP symmetry is possible. 
For example, additional scalar fields coupled to an order parameter can avoid symmetry restoration \cite{Dvali:1995cc,Dvali:1996zr,Baldes:2018nel,Glioti:2018roy,Meade:2018saz,Carena:2021onl,Murai:2024alz,Murai:2024bjy,Nakagawa:2025suc}.}

When the thermal mass becomes subdominant at $T\lesssim v_{\rm CP}$, $\eta$ develops a nonzero VEV of the order of $v_{\rm CP}$ which breaks CP symmetry.
In the next section, we will discuss the stabilization of the CP-violating phase.

\subsection{Backreaction on $S$ from $\eta$} \label{subsec:backreaction}

So far, we have ignored a backreaction effect on $S$ from $\eta$.
However, for the scaling solutions of $S$ to be effective, throughout the cosmic evolution, we generically require 
\begin{equation} \label{generalBR}
    \frac{\lambda_S^2}{(n!)^2 M_\text{Pl}^{2n-4}} \langle|S|\rangle^{2n} >\frac{\lambda}{m!} \frac{\langle |S|\rangle^{m} \langle|\eta|\rangle^\ell}{M_\text{Pl}^{m+\ell-4}},
\end{equation}
until $S$ starts to oscillate around its VEV at the origin. Note that during inflation, if $\langle\eta_\text{inf}\rangle =0$, this condition is trivially satisfied, which is the case for $\ell=2,3,4$ with properly chosen parameters, as we discussed in Sec.~\ref{eta_evolve}. However, if $\eta$ gets a large VEV during inflation, Eq.~\eqref{generalBR} gives a constraint on $\lambda$. Let us take $\ell=1$ as an example. Substituting Eq.~\eqref{Sinf} and Eq.~\eqref{etainf} into Eq.~\eqref{generalBR}, we have
\begin{equation}
    \lambda < \lambda|_\text{inf} \equiv m! \left( \frac{(4\lambda_\eta)^\frac{1}{3}\lambda_S^2}{(n!)^2} \right)^\frac{3}{4} \left( \frac{ \langle |S_\text{inf}| \rangle }{M_\text{Pl}}\right)^{-m+\frac{3n}{2}}.
\end{equation}
Similarly, for $\ell=1$, after the completion of reheating, we obtain\footnote{This analysis holds for $3n<2m+1$, which corresponds to the case that during the early radiation-domination epoch $|S|^m |\eta|^\ell$ decreases more rapidly than $|S|^{2n}$ as the temperature drops. In the opposite case,  $3n>2m+1$, $|S|^{2n}$  falls off faster, and one must ensure that it dominates until $T\gtrsim v_\text{CP}$.}
\begin{equation}
    \lambda < \lambda|_\text{RD} \equiv \lambda_S\frac{m!}{n!} \frac{T_\text{RH}}{M_\text{Pl}}\left( \frac{\langle |S_{\text{RD}}(T_\text{RH})|\rangle}{M_\text{Pl}}\right)^{n-m}.
\end{equation}
Combining these two conditions, the constraint reads 
\begin{equation} \label{lambdainfRD}
    \lambda< \min\{\lambda|_\text{inf}, \lambda|_\text{RD} \} \ .
\end{equation}

After $\eta$ obtains the VEV $\langle \eta \rangle \sim v_{\text{CP}}$, the backreaction
could modify the behavior of $S$ in Eq.~\eqref{SRD} if
\begin{equation} \label{backreactCond}
    \frac{\lambda_S^2}{(n!)^2M_{\text{Pl}}^{2n-4}}|S|^{2n} \sim \frac{\lambda}{m!\ell!M_{\text{Pl}}^{m+\ell-4}}|S|^m v_{\text{CP}}^{\ell} \ .
\end{equation}
For the case of $m>2n$, due to the temperature dependence of $|S_{\text{RD}}|\propto T^{\frac{2}{n-1}}$, the mixing term decreases faster than the self-interaction term of the $S$ field. To avoid the backreaction from the mixing term to change the dynamics of the $S$ field during the epoch from $|\eta|$ getting a VEV around $v_{\text{CP}}$ to the onset of the $S$ oscillation, we require
\begin{equation}
    \frac{\lambda_S^2\langle|S_{\rm RD}|\rangle^{2n}}{(n!)^2M_\text{Pl}^{2n-4}}\bigg|_{T\sim v_\text{CP}} > \frac{\lambda\langle|S_{\rm RD}|\rangle^mv_\text{CP}^\ell}{m! \ell!M_\text{Pl}^{m+\ell-4}}\bigg|_{T\sim v_\text{CP}}. 
    \label{backreaction}
\end{equation}
Substituting Eq.~\eqref{SRD} into this expression, one obtains the constraint,
\begin{align} \label{b_r_m_gtr_2n}
    \frac{\lambda(n!)^2}{\lambda_S^2 m!} \left[\frac{\pi^2g_*}{90} \frac{2(n-3)(n!)^2}{n(n-1)^2 \lambda_{{S}}^2} \right]^{\frac{m-2n}{2(n-1)}} \left(\frac{v_\text{CP}}{M_\text{Pl}} \right)^{\frac{2(m-2n)}{n-1}+\ell} <1 \ .
\end{align}

For $m=2n$, since both the self-interaction and mixing terms decrease with temperature at the same rate, the condition is independent of the temperature and given by
\begin{align} \label{b_r_m_eq_2n}
v_{\rm CP} \lesssim 2\times10^{13}\GeV \times \lmk\frac{(2n)!\ell!}{(n!)^2} \rmk^{\frac{1}{\ell}}\lmk\frac{\lambda}{10^{-1}}\rmk^{-\frac{1}{\ell}}\lmk\frac{\lambda_S}{10^{-3}}\rmk^{\frac{2}{\ell}}.
\end{align}

For $m<2n$, the mixing term decreases slower.
The backreaction from the mixing term can be neglected if $m_S$ is large enough for $S$ to start oscillating around the origin before Eq.~\eqref{backreactCond} is satisfied.
Then, we require
\begin{equation}
    \frac{\lambda_S^2\langle|S_{\rm RD}|\rangle^{2n}}{(n!)^2M_\text{Pl}^{2n-4}}\Bigg|_{H\sim m_S} > \frac{\lambda\langle|S_{\rm RD}|\rangle^mv_\text{CP}^\ell}{m!\ell! M_\text{Pl}^{m+\ell-4}}\Bigg|_{H\sim m_S},
\end{equation}
which leads to
\begin{align} \label{mSbackreaction}
    \lambda < &m!\ell!\left( \frac{\lambda_S}{n!}\right)^\frac{m-2}{n-1} \left( \frac{2(n-3)}{n(n-1)^2} \right)^\frac{2n-m}{2(n-1)} \notag \\
    &\ \times\left(\frac{M_\text{Pl}}{v_\text{CP}} \right)^\ell \left( \frac{m_S}{M_\text{Pl}}\right)^\frac{2n-m}{n-1}.
\end{align}
Although there exists an epoch that the mixing term can be larger than the self-interaction term, in most cases of $m>2$, $m_S$ dominates during this epoch so that the $S$ field oscillates around the origin.

\section{Phase mode dynamics} \label{subsec:eta}

Let us begin with the phase mode of $S$.
As discussed in \SEC{sec:evolve_S}, the scalar field $S$ has a large field value and its phase is randomly chosen as an initial condition during inflation.
Since inflation expands causal patches exponentially, a single value of the phase is assigned to our Universe.
The evolution of the phase mode depends on the VEV of $\eta$, because the mixing term gives the dominant potential.

First, consider the case of $\ell=2,3,4$ with proper parameters such that $\langle\eta\rangle=0$ during inflation.
Before the $\eta$ field acquires a nonzero VEV, the phase direction of $S$ is flat.
In this case, we can assume that the initial value is uniformly distributed, and is generically misaligned from the minimum of the potential from the mixing term.
As will be seen shortly, the phase mode of $S$ keeps constant even in the RD era, resulting in the domain wall collapse.

For $\ell=1$, $|\eta|$ gets a large VEV (\ref{etainf}) during inflation. 
Let us parameterize 
\begin{align}
    \eta_\text{inf}&=\frac{v_\eta^I}{\sqrt{2}} \exp\left[{i\lmk\theta^I_\eta+\frac{a}{v_\eta^I}\rmk}\right] \ , \notag \\[1ex]
    S_\text{inf}&=\frac{\chi^I}{\sqrt{2}} \exp{\left[i\lmk\theta^I_S+\frac{b}{\chi^I}\rmk\right]} \ ,
\end{align}
where $v_\eta^I\equiv\sqrt{2}\langle|\eta_\text{inf}|\rangle$ and $\chi^I \equiv \sqrt{2}\langle|S_\text{inf}|\rangle$, and $a$, $b$ are excitations along the phase directions whose VEVs are zero, $\langle a\rangle=\langle b\rangle=0$.
Then, the potential in phase directions can be extracted as
\begin{align} \label{Vphaseinf}
     V^{\inf}_\text{phase} 
     &\simeq \alpha (v_\eta^{I}) ^2\cos\left(2\theta_\eta^{I}+\frac{2a}{v_\eta^I}\right) + \frac{\beta (v_\eta^I)^4 }{2}\cos\left(4\theta_\eta^I+\frac{4a}{v_\eta^I}\right)  \notag \\
     &+ \frac{\lambda}{2^{\frac{m-1}{2}}m!} \frac{(\chi^I)^m v_\eta^I}{M_\text{Pl}^{m-3}} \cos\left[m\left(\theta_S^I+\frac{b}{\chi^I}\right)+\theta_\eta^I+\frac{a}{v_\eta^I}\right].
\end{align}
As we will discuss in detail later, when $T\lesssim v_\text{CP}$, to obtain a stabilized CP-violation vacuum, our setup prefers that $\alpha\sim \beta v_\text{CP}^2$.
Thus, at the vacuum, $\beta (v_\eta^I)^4$ is dominant compared to $\alpha (v_\eta^I)^2$, and $\theta_\eta^I$ is stabilized at $\pi/4$ during inflation.
To minimize the mixing term, we obtain
\begin{equation} \label{thetaSinf}
    \theta^I_S = \frac{\pi}{m}\lmk(2k+1)-\frac{1}{4}\rmk ,
\end{equation}
with $k\in \mathbb{Z}$.
Furthermore, Eq.~\eqref{Vphaseinf} also determines the masses of $a$ and $b$ modes during  inflation:
\begin{align}
 & m_a^2 = \frac{\partial^2 V^{\inf}_\text{phase}}{\partial a^2}\bigg|_{a,b\rightarrow0} \simeq\frac{\sqrt{2}\lambda }{ m!}\left(\frac{\langle S_\text{inf} \rangle}{M_\text{Pl}}  \right)^m \frac{M_{\text{Pl}}^3}{ v_\eta^I} \ , \notag \\[1ex]
  &m_b^2=\frac{\partial^2 V^{\inf}_\text{phase}}{\partial b^2}\bigg|_{a,b\rightarrow0} \simeq  \frac{\lambda m^2}{\sqrt{2} m!} \left(\frac{\langle S_\text{inf} \rangle}{M_\text{Pl}}  \right)^{m-2} v_\eta^I M_\text{Pl} \ .
\end{align}
In our focused parameter region that will be clarified in \SEC{sec_DM}, $m_a$ is generally larger than the Hubble scale of the inflationary epoch, while the competition between $m_b$ and $H_\text{inf}$ depends on the parameter choices. It indicates that there are two different scenarios: for $m_b\gtrsim H_\text{inf}$ during inflation, $b$ can roll down to its stabilized VEV and $\theta_S$ is given by Eq.~\eqref{thetaSinf}. On the other hand, if $m_b<H_\text{inf}$, then $b$ cannot move during inflation, and $\langle S_\text{inf} \rangle$ carries its randomized initial phase during this epoch. The boundary value of $\lambda$ between these two scenarios is given by
\begin{equation} \label{bmove}
    \lambda|_{b, \rm move} = \frac{m!(4\lambda_\eta)^{\frac{1}{4}}}{m^{\frac{3}{2}}} \left( \sqrt{\frac{c_S}{n}} \frac{n!}{\lambda_S} \right)^{\frac{3-2m}{2(n-1)}} \left(\frac{H_\text{inf}}{M_\text{Pl}} \right)^{\frac{3}{2}-\frac{2m-3}{2(n-1)}}.
\end{equation}

After reheating, when $T\gtrsim v_\text{CP}$, the $\eta$'s VEV is determined by Eq.~\eqref{thermalETAVEV}, which is around the origin. 
Since the mass of $b$ is negligible compared to the Hubble parameter, $\theta_S$ maintains its value
during inflation,
which provides a CP violation source that results in the domain wall collapse, as we will see in the later discussion.

Next, let us study the stabilization of the CP-violating phase of $\eta$ at $T\lesssim v_{\rm CP}$.
Parameterizing $\eta$ as $\eta=|\eta| \, e^{i\theta}$,
we obtain the potential relevant to the stabilization of $\eta$,
\begin{equation}
    V_{\eta} = \lambda_\eta \left( |\eta|^2-\frac{v_{\text{CP}}^2}{2} \right)^2 +2\alpha|\eta|^2\cos{2\theta} + 2\beta|\eta|^4\cos{4\theta} \ .
\end{equation}
Note that the $\eta-S$ mixing term is irrelevant to the stabilization, because it is Planck-suppressed and subdominant.
By solving $\del V_\eta/\del|\eta|=\del V_\eta/\del\theta=0$, we find the VEVs of $|\eta|$ and $\theta$,
\begin{align} \label{veta}
    & \frac{v_\eta^2}{2} \equiv \langle|\eta|\rangle^2 = \frac{v_{\text{CP}}^2}{2(1-2\beta/\lambda_\eta)} \ ,\\[1ex]
& \theta_\eta \equiv \langle\theta \rangle = \frac{1}{2} \cos^{-1} \lmk-\frac{\alpha}{2\beta v_\eta^2}\rmk.
\end{align}
These solutions satisfy the following conditions:
\begin{align}
    & 2\beta/\lambda_\eta<1 \ ,\label{bounded}\\[1ex]
    &|\alpha| \sim 2\beta v_\eta^2 \ , \quad \beta>0 \ . 
\end{align}
The former condition is required for the potential to be bounded from below, while the latter condition is that a sizable CP violation is generated.
If $\beta<0$, all possible vacua would be CP-conserving. 

\begin{figure}
    \centering
    \includegraphics[width=0.99\linewidth]{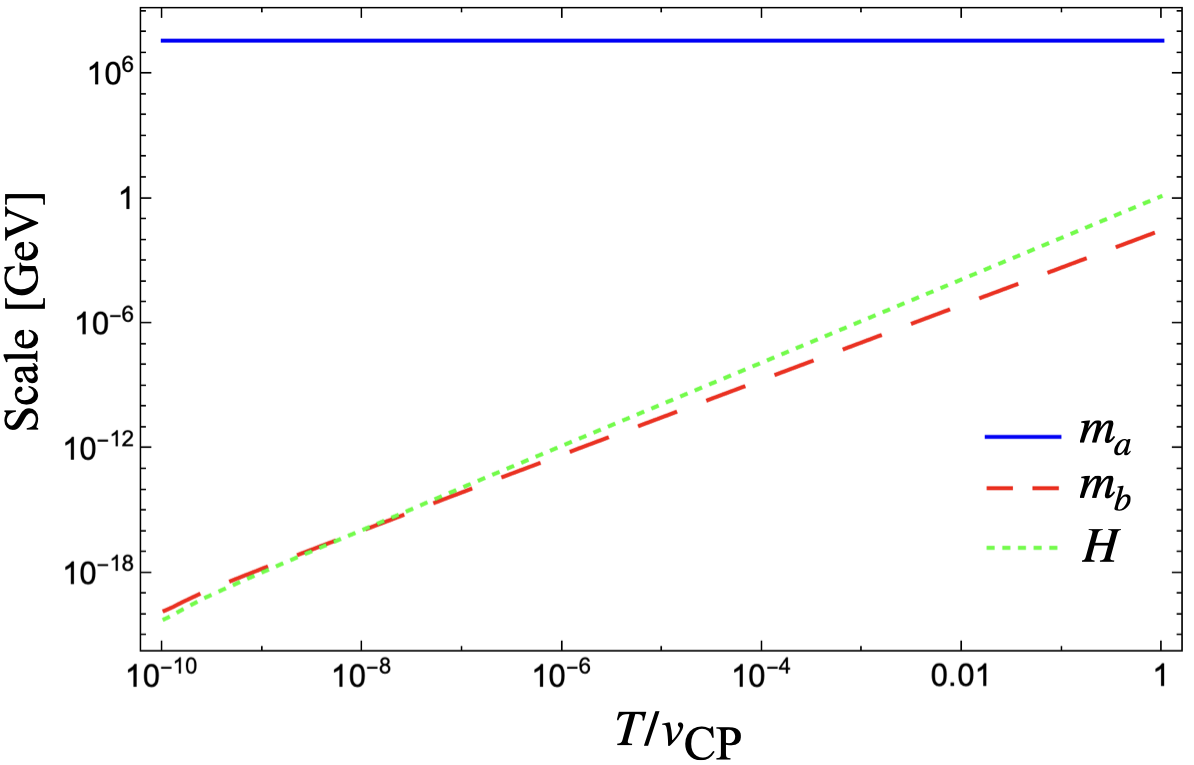}
    \caption{The temperature dependence of $m_a$, $m_b$ and $H$ for $(\ell,m,n) = (1, 11,6)$.
    Here we set $v_{\text{CP}}=10^{9}$ GeV, $\alpha=v_{\text{CP}}^2\beta$, $\beta = 10^{-2}$, $\lambda=10^{-1}$, $\lambda_\eta=1$, and $\lambda_S =10^{-3}$.}
    \label{fig:maVSmbVSH}
\end{figure}

Focusing on the axial directions, we expand the fields after the CP symmetry breaking around $\theta_\eta$ and $\theta_S$, respectively, as
\begin{align}
\eta &\simeq \frac{v_{\eta}}{\sqrt{2}} \exp \left[i\lmk\theta_{\eta} +\frac{a}{v_{\eta}}\rmk\right] \ , \\[1ex] 
S&\simeq \frac{\chi}{\sqrt{2}} \exp \left[i\lmk\theta_S+\frac{b}{\chi}\rmk\right] \ ,
\end{align}
where $\theta_S=\theta_S^I$ is given by Eq.~\eqref{thetaSinf}, for $\lambda \gtrsim \lambda|_{b, \rm move}$, or is randomized for the opposite case.
Substituting them into the $\eta-S$ mixing term, we obtain 
\begin{equation} \label{Vbias}
    V_{\text{mix}} \simeq  \frac{1}{\ell^2} \Lambda_{\cancel{\text{CP}}}^2 v_{\eta}^2\cos\left[\ell\lmk\frac{a}{v_{\eta}}+\theta_\eta\rmk+m\lmk\frac{b}{\chi}+\theta_S\rmk\right] \ .
\end{equation}
Here $\Lambda_{\cancel{\text{CP}}}^2$ denotes a contribution to the mass-squared of the $a$ mode from this potential term, defined as
\begin{equation} \label{LambdaCP}
\Lambda^2_{\cancel{\text{CP}}} =  \frac{\lambda \ell^2}{2^{\ell/2-1} m!\ell!} \frac{\langle|S_{\text{RD}}|\rangle^m v_{\eta}^{\ell-2}}{M_{\text{Pl}}^{m+\ell-4}} \ .
\end{equation}
Compared with the CP breaking scale, $\Lambda_{\cancel{\rm CP}}$ is hierarchically small.
At $T=v_{\rm CP}$, the ratio is estimated as
\begin{align}
\frac{\alpha}{\Lambda_{\cancel{\rm CP}}^2} \sim 1.2 \times 10^7 \lmk\frac{\beta/\lambda}{10^{-1}}\rmk\lmk1-\frac{2\beta}{\lambda_\eta}\rmk^{-\frac{1}{2}} \lmk\frac{v_{\rm CP}}{10^9\GeV}\rmk^{-\frac{7}{5}}.
\end{align}
Here we set $(\ell, m,n) = (1,11,6)$, $\alpha=\beta v_{\rm CP}^2$, $\lambda_S=10^{-3}$, and $g_*=106.75$ as reference values. 
Although the hierarchy depends on the parameter choice, typically we have $\alpha\gg \Lambda_{\cancel{\rm CP}}^2$.

Including the potential (\ref{Veta}), we can write down the full potential relevant to the phase mode excitations $a$ and $b$ up to a constant,
\begin{align}\label{Vab}
    V(a,b)
    =\, &V_{\rm mix}+\alpha v_{\eta}^2\cos(2\theta_{\eta}+2a/v_{\eta}) \notag \\
    &+\frac{\beta v_{\eta}^4}{2} \cos(4\theta_{\eta}+4a/v_{\eta}) \ .
\end{align}
The minimum of the $a$ mode is almost unmodified, $\langle a\rangle\approx0$, because $\alpha\gg\Lambda_{\cancel{\rm CP}}^2$. 
On the other hand, the true minimum of the $b$ mode after the CP breaking is estimated as the solution to $\del V(a,b)/\del b = 0$, which is given by
\beq
\langle b \rangle = -\frac{\chi}{m}(\ell\theta_\eta+m\theta_S + (2k+1)\pi), \ \  k\in \mathbb{Z} \ .
\eeq
Around $a=\langle a\rangle$ and $b=\langle b\rangle$, we respectively obtain the mass terms for $a$ and $b$ after SCPV as
\begin{align} \label{m_a}
    m_a^2 &= \frac{\partial^2 V(a,b)}{\partial a^2} \bigg|_{a\rightarrow0, \ b \rightarrow \langle b \rangle} 
    = 8\beta v_{\eta}^2-\frac{2\alpha^2}{\beta v_{\eta}^2}+\Lambda_{\cancel{\text{CP}}}^2 \ ,
\end{align}
and
\begin{align}
    m_b^2 &= \frac{\partial^2 V(a,b)}{\partial b^2} \bigg|_{a\rightarrow0, \ b \rightarrow \langle b \rangle}
    = \frac{m^2}{\ell^2}\frac{v_{\eta}^2}{\chi^2}\Lambda_{\cancel{\text{CP}}}^2 \ .
\end{align}
Without cancellation, the first two terms in $m_a^2$ roughly give $v_{\text{CP}}^2$.
Since the mass of $b$ originates from the mixing term, $b$ is much lighter than $a$.

\FIG{fig:maVSmbVSH} shows the temperature dependence of $m_a$ and $m_b$ as well as the Hubble parameter $H$.
We can see that $m_b$ is smaller than $H$ in the range of not too small $T/v_{\rm CP}$.
This corresponds to the range where the backreaction can be neglected.
To see this, dividing both sides of \EQ{backreactCond} by $|S|^2$, we obtain
\beq
\frac{|S_{\rm RD}|^{2n-2}}{\Mpl^{2n-4}} \sim \frac{|S_{\rm RD}|^{m-2}v_{\rm CP}^\ell}{\Mpl^{m+\ell-4}} \ ,
\eeq
where we ignore the coefficients for simplicity.
The left-hand side is $H^2$ in terms of \EQ{SRD}, while the right-hand side is the same as $m_b^2$.
Thus the phase mode $b$ does not move, as long as the backreaction from the mixing term is inefficient.\footnote{This is shown explicitly by some arithmetic.
Let us define $T_{\rm BR}$ as the temperature at which the backreaction becomes efficient and $T_{{\rm osc},b}$
as the temperature at which $b$ starts to oscillate.
We can find $T_{\rm BR}/T_{{\rm osc},b} = [2^{\ell/2-1}n(n-1)^2/m^2(n-3)]^{(n-1)/(4n-2m)}$, which is $\mathcal{O}(1)$, independent of the choice of $(\ell,m,n)$.}
The dynamics of $S$ at later times will be described in \SEC{sec_DM}.

\section{CP domain wall decay}
\label{domain_wall}

\begin{figure}[t!]
\centering
\includegraphics[width=0.99\linewidth]{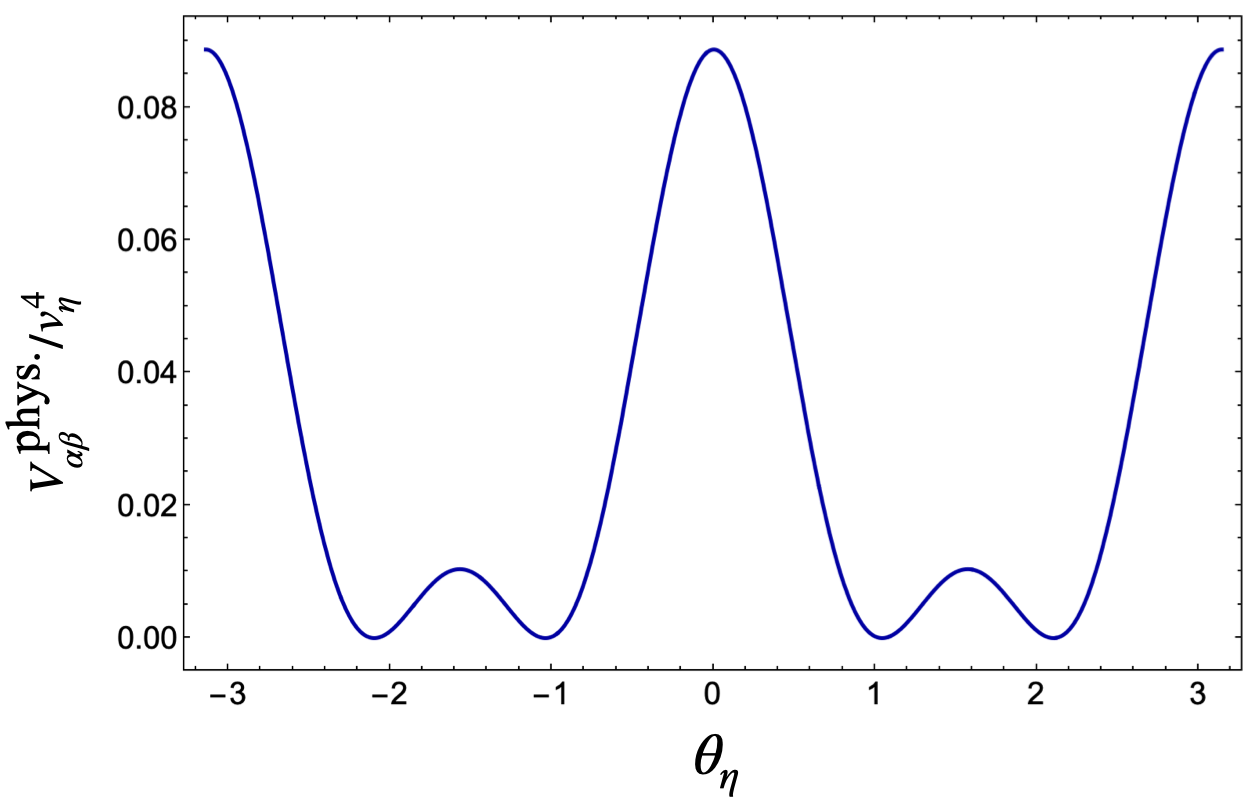}
\caption{
The shape of the potential for $\theta_\eta$.
Here we define $V^{\text{phys.}}_{\alpha \beta}\equiv V_{\alpha \beta}- \min\{V_{\alpha \beta}\}$ with $V_{\alpha \beta}(\theta_\eta) = \alpha v_{\eta}^2\cos(2\theta_{\eta})
+(\beta v_{\eta}^4/2) \cos(4\theta_{\eta})$.
We take $\beta=10^{-2}$, $\alpha=\beta v_\text{CP}^2$, and $\lambda_\eta=1$.
}
\label{fig:DWProfile}
\end{figure}

As the temperature of the Universe cools down to the CP breaking scale, $T \sim v_{\text{CP}}$, $\eta$ gets a complex VEV with a stabilized phase, spontaneously breaking CP symmetry and the $Z_2$ symmetry, under which both $\eta$ and $S$ are odd, forming CP domain walls due to the equivalence of $\theta_\eta$ and $-\theta_\eta$, and $Z_2$ domain walls due to the equivalence of $\theta_\eta$ and $\theta_\eta \pm \pi$. The shape of the potential for $\theta_\eta$ is shown in Fig.~\ref{fig:DWProfile}. The four minima correspond to the spontaneous breaking of both CP and $Z_2$ symmetries.
However, since the $S$ field has a large VEV $\langle S_{\rm RD} \rangle$ with a complex phase $\theta_S$, $\eta$ feels an effective CP-violating perturbation through the $\eta-S$ mixing term in Eq.~\eqref{Vtot}, which behaves as a bias term to drive the CP and $Z_2$ domain walls to decay.

Let us discuss the process of the domain wall collapse. We consider the potential for the phase directions at the minima, 
\begin{align}
   V(\theta_\eta, \theta_S) &= V(a,b)|_{a,b\rightarrow0}  \notag \\
   &= \frac{1}{\ell^2} \Lambda_{\cancel{\text{CP}}}^2 v_{\eta}^2\cos(\ell\theta_\eta+m\theta_S) \notag \\ 
   &\ \ \ \ + \alpha v^2_{\eta} \cos(2\theta_\eta) + \beta \frac{v^4_{\eta}}{2} \cos(4 \theta_\eta) \ .
\end{align}
The bias term lifts the degeneracy among the vacua and generates a pressure difference across the domain wall. To determine whether the domain wall network collapses, it is crucial to compare the volume pressure induced by the bias term with the tension force of the domain wall. We now estimate these two effects in turn. 
The volume pressure $p_V \sim \Delta V_{\text{bias}}$ is determined by the first term of the above potential,
\beq
p_V \simeq \frac{\Lambda^2_{\cancel{CP}} v_\eta^2}{\ell^2} \ ,
\eeq
which is proportional to $H^{m/(n-1)}$.
On the other hand, assuming that the wall network enters into the scaling regime immediately, the tension force is expressed as
\begin{equation} \label{pT}
    p_T \simeq H \int^{(\theta_\eta)_{\rm min2}}_{(\theta_\eta)_{\rm min1}} d\theta_{\eta} v_{\eta}\sqrt{V^{\rm phys.}_{\alpha\beta}(\theta_\eta)} \ ,
\end{equation}
in which $(\theta_\eta)_{\rm min1,2}$ are the adjacent vacua, and
$V^{\text{phys.}}_{\alpha \beta}\equiv V_{\alpha \beta} - \min\{V_{\alpha \beta}\}$, so that the square root is well-defined, and here $V_{\alpha \beta}(\theta_\eta) \equiv \alpha v_{\eta}^2\cos(2\theta_{\eta})+(\beta v_{\eta}^4/2) \cos(4\theta_{\eta})$.
The integral determining the tension force can be evaluated analytically, yielding
\begin{align}
    p_T &\simeq 2\sqrt{\beta}v_\eta^3 H \notag \\
    &\ \ \ \ \times \left[\sqrt{1-\left(\frac{\alpha}{2\beta v_\eta^2}\right)^2}  -\frac{\alpha}{2\beta v_\eta^2} \arcsin{\left(\frac{\alpha}{2\beta v_\eta^2}\right)} \right] \notag \\
    &\sim  \mathcal{O}(1) \times  \sqrt{\beta} v_\eta^3H \ .
\end{align}
The dimensionless bracket is generically of order unity throughout the parameter region of interest. Therefore, the only explicit temperature dependence arises from the Hubble parameter, $H\propto T^2$.

When the volume pressure is comparable to the tension force, the system tends to destabilize. Hence, we need to clarify how the tension force $p_T$ and the volume pressure $p_V$ evolve as the temperature of the Universe decreases.

\begin{figure}
    \centering
    \includegraphics[width=0.99\linewidth]{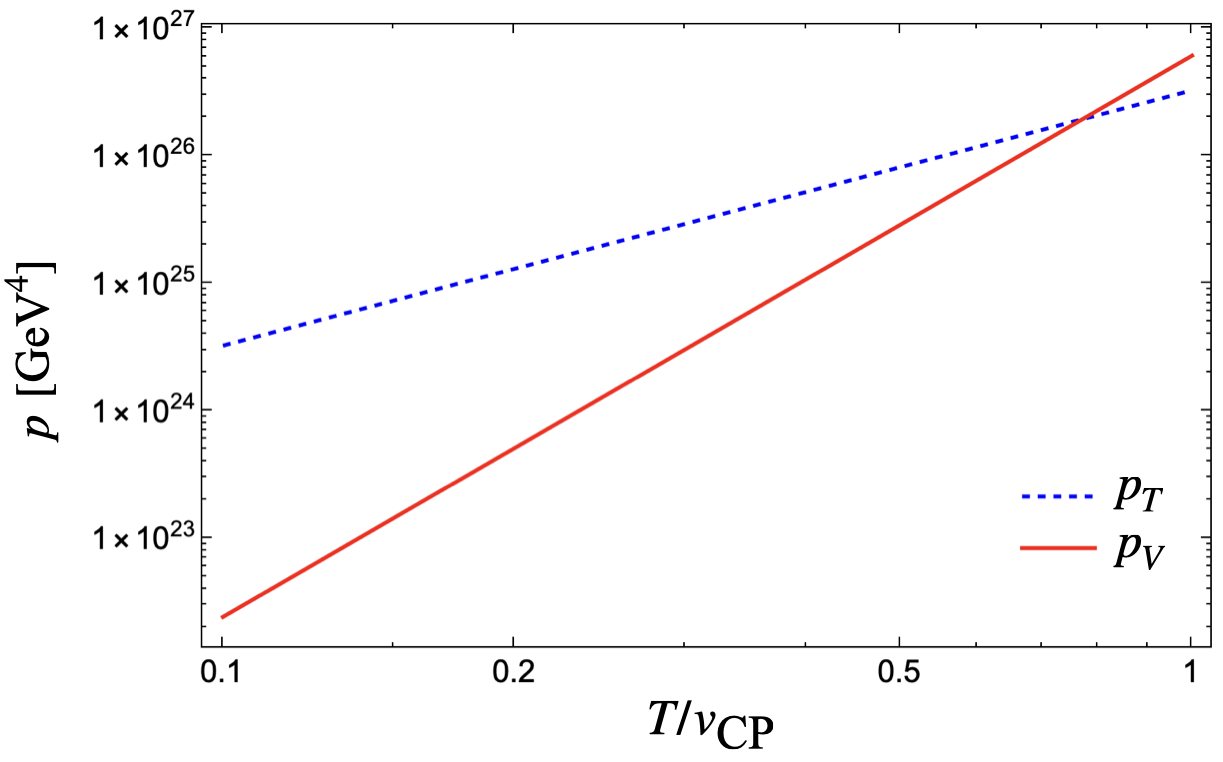}
    \caption{The evolution of the volume pressure $p_V$ (red solid) and the tension force $p_T$ (blue dashed) with respect to the temperature $T$. 
    The parameters are chosen as $(\ell,m,n) = (1,11,6)$, $\lambda_\eta =1$, $\lambda=10^{-1}$, $\lambda_S=10^{-3}$, $\beta=10^{-2}$, $\alpha=\beta v_{\text{CP}}^2$, and $v_{\text{CP}} = 10^{9}$ GeV.
    }
    \label{fig:pVT}
\end{figure}

We first discuss the case of $m \geq n-1$ where the volume pressure falls off more rapidly than, or at the same rate as, the tension force.
At the temperature of the domain wall formation $T \sim v_{\text{CP}}$, we require $p_V >p_T$, so that the domain walls decay sufficiently fast.\footnote{In the presence of a large explicit breaking, domain walls cannot form due to the non-uniform initial distribution of scalar fields \cite{Coulson:1995nv,Larsson:1996sp}.
It can be realized if the probability $P_t$ of being assigned to the true minimum is large enough compared to that to the false minimum $P_f$.
This is not our case, because of $P_f/P_t\sim \exp(-\Delta V_{\rm bias}/|V_{\alpha\beta}|)\sim \exp(-\Lambda_{\cancel{\rm CP}}^2/\alpha) \sim 1$.}
Fig.~\ref{fig:pVT} shows the temperature dependence of $p_V$ and $p_T$.

Next, we consider the less straightforward case, $m < n-1$.
In this case, the volume pressure decreases slower than the tension force. It is sufficient to require $p_V > p_T$ at $T \sim v_{\text{CP}}$. However, even when $p_V < p_T$ at $T \sim v_{\text{CP}}$, since the tension decreases faster, there exists a temperature $T_c$ such that $p_V(T_c) = p_T(T_c)$. Such $T_c$ must be larger than the temperature that the field $S$ starts to oscillate, $T_{\rm osc}$, otherwise the bias term would be ineffective before $T=T_c$. In order to solve the domain wall problem for $m<n-1$, we then require\footnote{Note that the case of $m<n-1$ is always included in the regime of $m<2n$ relevant for the backreaction analysis.
The bound of \EQ{mSbackreaction} requires either an extremely small $\lambda$ or an exceptionally low CP-breaking scale $v_{\rm CP}$, neither of which is well motivated.
We therefore do not consider this scenario further.}
\begin{equation}
    p_V(T_{\rm osc}) > p_T(T_{\rm osc}) \ ,
\end{equation}
in which $T_{\rm osc}$ is defined by $H(T_{\rm osc}) \simeq m_S$.

\section{Dark matter}
\label{sec_DM}

The scalar field $S$ obeys the scaling solution until $H\lesssim m_{{S}}$, after which it starts to oscillate around its vacuum expectation value, $\langle S \rangle = 0$.
This coherent oscillation is regarded as cold DM.
\footnote{In this paper, we focus on the case that the backreaction is inefficient, as discussed in \EQ{backreaction}.
Once the backreaction becomes significant before $H\simeq m_S$, the phase mode $b$ starts to move, while the amplitude $|S|$ becomes constant, because the mixing term becomes negative at the minimum of the potential in the phase direction, which was discussed in Ref.~\cite{Hao:2025kcz}.
In this case, $b$ can be a DM candidate instead of $|S|$.
}
The initial oscillation amplitude is given by
\beq
\chi_{\rm osc} = \sqrt{2}|\langle S_{\rm RD}(T_{\rm osc})\rangle| \ . 
\eeq
Using the entropy density $s(T)=(2\pi^2/45)g_{*s}T^3$ with $g_{*s}$ the effective number of relativistic degrees of freedom for entropy, we estimate the adiabatic invariant,
\begin{align}
\left.\frac{n_\chi}{s}\right|_{T=T_{\rm osc}} = \frac{1}{2}m_S \chi^2_{\rm osc}/s(T_{\rm osc}) \ ,
\end{align}
where $n_\chi$ represents the number density of $\chi$.
Thus, the abundance of $S$ is evaluated as
\begin{align}
    \Omega_\chi h^2 &= \left.\frac{m_S n_\chi}{s}\right|_{T=T_{\rm osc}} \frac{s(T_0)}{\rho_c h^{-2}} \notag\\[1ex]
    &\simeq 0.12 \left(\frac{\lambda_S}{10^{-3}}\right)^{-\frac{2}{5}}\left(\frac{m_S}{1.6\times10^{-14} \GeV}\right)^{\frac{9}{10}},
\end{align}
where $\rho_c$ denotes the critical density, $T_0$ is the current temperature of the Universe, and $h$ is the reduced Hubble constant.
We take $(\ell,m,n) = (1,11,6)$, and $g_{*s}(T_{\rm osc}) \simeq g_*(T_{\rm osc}) = 80$.

\begin{figure*}[htbp]
  \centering
  \begin{minipage}[t]{0.45\textwidth}
    \centering
    \includegraphics[width=\textwidth]{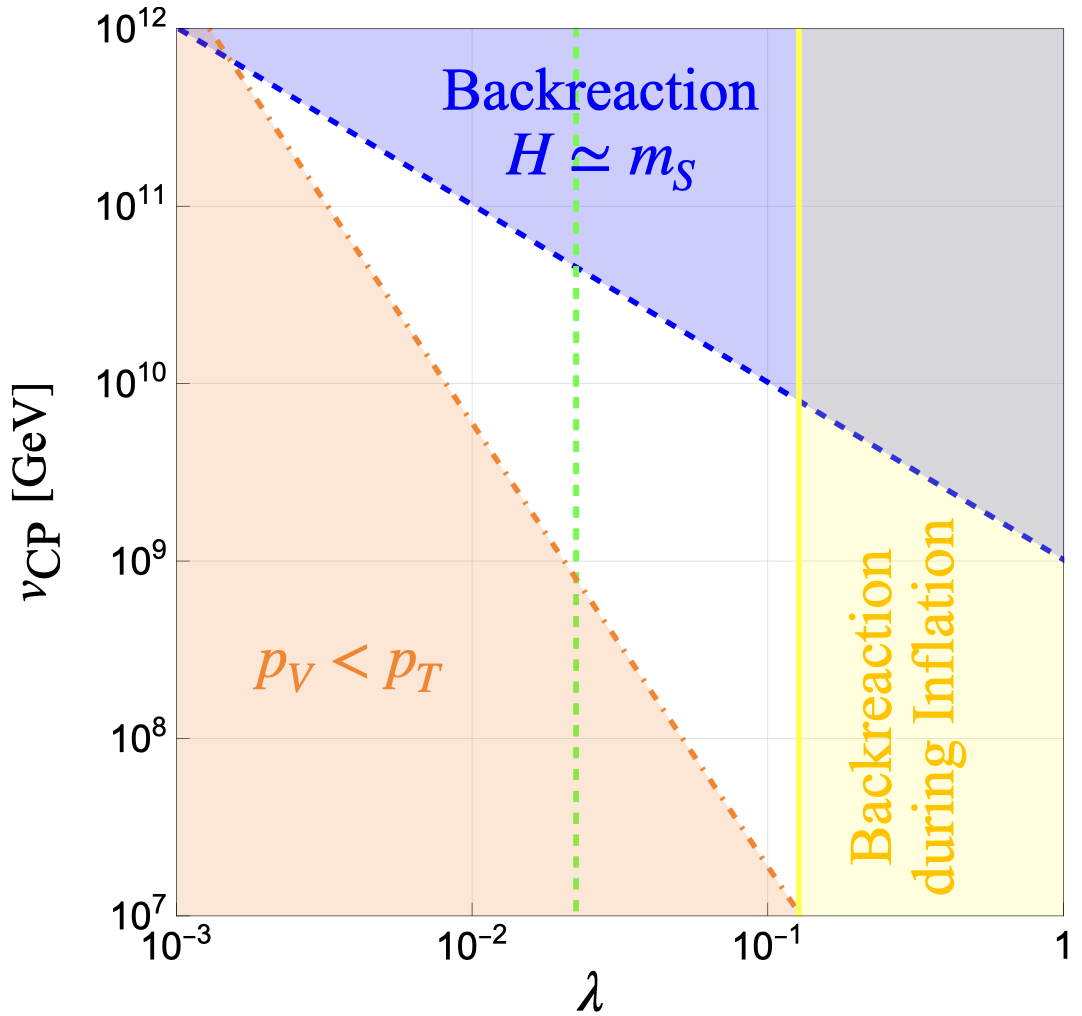}
  \end{minipage}%
  \hspace{0.05\textwidth}
  \begin{minipage}[t]{0.45\textwidth}
    \centering
    \includegraphics[width=\textwidth]{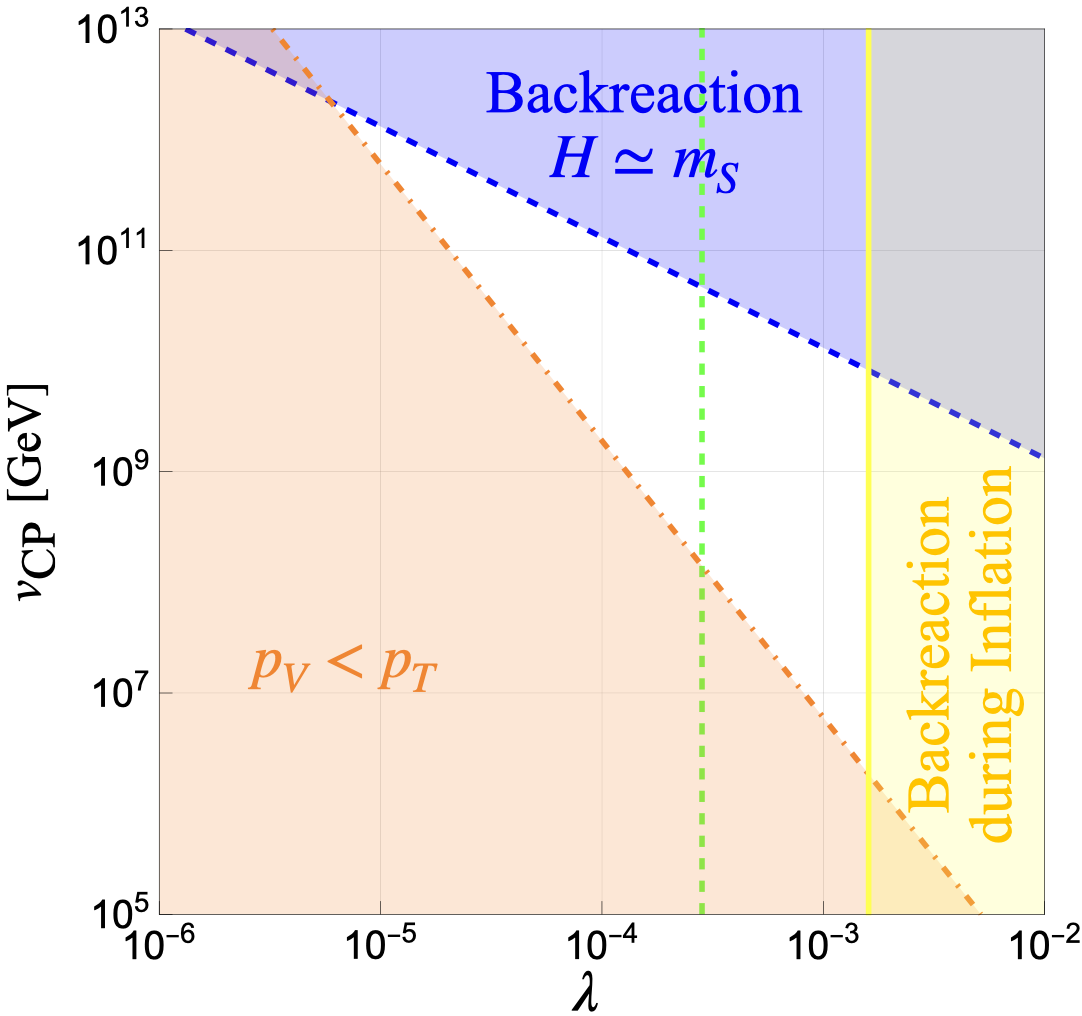}
  \end{minipage}
  \caption{The viable parameter space for $\ell=1$, $n=6$, and $m=11$, $\lambda_\eta =1$, $\beta=10^{-2}$, and $\alpha=\beta v_\text{CP}^2$.
  In addition, $m_S$ is fixed by the requirement $\Omega_\chi = \Omega_\text{DM}$ for the given $\lambda_S$.
  The left and right panels correspond to $\lambda_S=10^{-3}$ and $\lambda_S=10^{-4}$, respectively. 
  The white region gives a viable parameter space.
  The orange region is excluded since $p_V<p_T$, implying that the domain walls do not decay.  
  The yellow region is excluded by Eq.~\eqref{lambdainfRD}, because the backreaction is non-negligible during inflation. 
  The blue region is excluded because the condition in Eq.~\eqref{mSbackreaction} is violated; the backreaction controls the evolution of $|S|$ before its coherent oscillation.  
  The vertical green dashed line corresponds to Eq.~\eqref{bmove}.  
  On the left-hand side of this line, the initial phase $\theta_S$ is randomized, while on the right it is set as $\theta_S$ determined by Eq.~\eqref{thetaSinf}.}
  \label{fig:ParamSpace}
\end{figure*}

Summarizing the discussion we have made so far, we can show which parameters are required for the sizable CP violation and the observed DM abundance.
The viable parameter space is displayed as the white region in \FIG{fig:ParamSpace}.
We fix  $\ell=1$, $n=6$, $m=11$, $\lambda_\eta =1$, $\beta=10^{-2}$, and $\alpha=\beta v_\text{CP}^2$.
In addition, $m_S$ is fixed by the requirement $\Omega_\chi = \Omega_\text{DM}$ for the given $\lambda_S$.
The left and right panels correspond to $\lambda_S=10^{-3}$ and $\lambda_S=10^{-4}$, respectively. The yellow region is excluded by Eq.~\eqref{lambdainfRD}, as the backreaction dominates during inflation. The orange region is excluded since $p_V<p_T$, implying that the domain walls do not decay.  The blue region is excluded because the condition in Eq.~\eqref{mSbackreaction} is violated; there the backreaction controls the evolution of $|S|$ before its coherent oscillation. The vertical green dashed line corresponds to Eq.~\eqref{bmove}.  On the left of this line, the initial phase $\theta_S$ is randomized, while on the right it is set as $\theta_S$ determined by Eq.~\eqref{thetaSinf}.

\section{UV completion}
\label{sec_UV}

The form of the potential \eqref{Vtot} is highly non-generic. In particular, it assumes the absence of lower-dimensional operators involving $|S|$ as well as $\eta-S$ mixing terms with powers different from those appearing in our setup. To justify this structure without fine tuning, we investigate a supersymmetric UV completion of the model.
Here, let us focus on the case of $n=6$, $m=19$ and $\ell=1$ as an example and consider the following superpotential (for simplicity, all dimensionless couplings are set to unity):
\begin{align}
\label{superpo}
    W &= X(\eta \bar\eta-v_\text{CP}^2)+\frac{1}{M_\text{Pl}^4}YS^6 \notag \\ 
     & \ \ \ +Z \left(\frac{1}{\M_\text{Pl}^5}S^6\eta+\frac{1}{M_\text{Pl}^{11}}S^{13} \right)+ U\bar\eta^2+ \frac{1}{M_\text{Pl}^6} X\bar\eta S^7 ,
\end{align}
where $\eta$, $\bar{\eta}$, $S$, $X$, $Y$, $Z$ and $U$ all denote chiral superfields. We impose a $U(1)$ symmetry and an R-symmetry to constrain the superpotential form. The charge assignment of each chiral superfield is summarized in TABLE~\ref{Tab_1}. Additionally, there exists an explicit $U(1)$ breaking term, with relevant dimensionality,
\begin{align}
\label{deltaW}
   \Delta W = v_\text{CP}U(\eta+\bar{\eta}) \ ,
\end{align}
where we assume that the parameter with unit mass dimension is at the CP breaking scale $v_\text{CP}$.

The superpotential terms in Eqs.~\eqref{superpo}, \eqref{deltaW} give the scalar potential terms relevant for the dynamics of $\eta, \bar \eta$,
\begin{align} \label{VetabaretaS}
    V(\eta,\bar\eta,S) &\supset \left| \frac{\partial W}{\partial X}\right|^2 + \left| \frac{\partial W}{\partial U} \right|^2 \notag \\
    &\simeq |\eta\bar\eta-v_\text{CP}|^2 + \left\{(\eta\bar\eta-v_\text{CP}^2)\frac{\bar\eta^*S^{*7}}{M_\text{Pl}^6}+{\rm h.c.} \right\} \notag \\
    &\ \ \ + \frac{|\bar\eta|^2|S|^{14}}{M_\text{Pl}^{12}} + |\bar\eta^2+v_\text{CP}(\bar\eta+\eta)|^2 \ .
\end{align}
Focusing on the terms that are not Planck-suppressed, we have
\begin{align}
    V(\eta, \bar\eta) \simeq & \, |\eta\bar\eta-v_\text{CP}|^2 + |\bar\eta|^4 + \left\{ v_\text{CP}(\bar\eta+\eta)\bar{\eta}^{*2}+ {\rm h.c.} \right\} \notag \\[1ex] &+v_\text{CP}^2 \left\{ |\bar\eta|^2+|\eta|^2+ \left(\bar\eta\eta^*+ {\rm h.c.} \right) \right\}  .
\end{align}
We can check that this potential stabilizes $\langle\eta\rangle= v_1e^{i\theta_1}$ and $\langle\bar\eta\rangle= v_2e^{i\theta_2}$ with non-trivial $\theta_{1,2}$.

\renewcommand{\arraystretch}{1.5} 
\begin{table}[!t] 
\begin{tabular}{|c|c|c|c|c|c|c|c|c|}
  \hline
    & \ \ $X$ \ \ & \ \ $Y$ \ \ & \ \ $Z$ \ \ & \ \ $U$ \ \ & \ \ $S$ \ \ & \ \ $\eta$ \ \ & \ \ $\bar{\eta}$ \ \ \\
    \hline
   \  $U(1)$ \  & 0 & $-6$ & $-13$ & 14 & 1 & 7 & $-7$ \\
    \hline
   \  $U(1)_R$ \  & 2 & 2 & 2 & 2& 0 & 0 & 0 \\
     \hline
\end{tabular}
\centering
\vspace{0.2cm}
\caption{The charge assignments of chiral superfields in the SUSY realization.}
\label{Tab_1}
\end{table}

Note that the Planck-suppressed operators
in Eq.~\eqref{VetabaretaS} are harmless. First, for the term,
$$(\eta\bar\eta-v_\text{CP}^2)\frac{\bar\eta^*S^{*7}}{M_\text{Pl}^6}+ \rm h.c. \ ,$$
when $T \gtrsim v_\text{CP}$, $\eta$ and $\bar\eta$ are settled around the origin due to thermal mass terms, so that it does not affect the dynamics of $S$. In addition, after $\langle \eta\bar\eta \rangle$ gets a nonzero VEV around $v_\text{CP}^2$,
this term becomes negligible. 
Moreover, the other term, $\frac{|\bar\eta|^2|S|^{14}}{M_\text{Pl}^{12}}$,
gives a sub-leading backreaction effect on the VEV of $S$.

The dynamics of the $S$ field is determined by
\begin{align}
    V(S)&\supset \left|\frac{\partial W}{\partial Y}\right|^2+m_S^2|S|^2 \notag \\[1ex]
    &= \frac{|S|^{12}}{M_\text{Pl}^8}+m_S^2|S|^2 \ ,
\end{align}
corresponding to the case with $n=6$.
The second term denotes the soft SUSY breaking mass-squared of $S$. We typically assume
$m_S \sim 10^{-15}\,\mathrm{GeV}$, which may require a severe fine tuning.
Generally, the mass of $S$ is expected to be at least of
the order of the gravitino mass. While the gravitino mass may be
considerably smaller than the TeV scale, obtaining
$m_S \sim 10^{-15}\,\mathrm{GeV}$ still requires a tuning.
As discussed in Ref.~\cite{Hao:2025kcz}, one possible way to alleviate this issue is to sequester the SUSY-breaking
sector from the $S$ sector, for instance through a no-scale K\"ahler
structure~\cite{Goto:1991gq}. In such a setup, the fermionic partner of $S$ can remain exceptionally light.
Understanding its cosmological consequences is a nontrivial problem, which we leave for a future investigation.

The mixing term that drives the domain wall decay is given by
\begin{align}
    V_\text{mix} (\eta,S) &\supset \left| \frac{\partial W}{\partial Z} \right|^2 \notag \\[1ex]
    &=\frac{|\eta|^2|S|^{12}}{M_\text{Pl}^{10}}+\frac{|S|^{26}}{M_\text{Pl}^{22}}+\frac{\eta|S|^{12}S^{*7}}{M_\text{Pl}^{16}}+ \rm h.c. 
\end{align}
The first term is relevant for the backreaction effect while the third term plays a role in the domain wall decay.
The detailed analysis in this case is left for a future study.

\section{Conclusion and Discussion}
\label{sec:Discussion}

We have explored a new dynamical mechanism to eliminate CP domain walls without introducing a permanent explicit CP violation. A  light scalar field carrying a large field value in the early Universe induces a transient bias among degenerate CP vacua, generating a volume pressure that can overcome the domain wall tension and trigger the collapse of the domain wall network. We derived conditions for successful domain wall decay, identified the viable parameter space, and showed that the same scalar field can naturally account for the observed DM abundance through coherent oscillations. Our mechanism can be applied to a broad class of spontaneous CP violation models and provides a new connection between the strong CP problem and DM.

Recently, the authors of Ref.~\cite{McNamara:2022lrw} sharpened the CP domain wall problem by arguing that CP domain walls are topologically protected when CP is realized as a gauged spacetime symmetry, as expected in quantum gravity. In such a case, an explicit CP-breaking operator is not available, since explicit breaking of a gauged symmetry is inconsistent. Our mechanism differs from this approach in that the fundamental theory preserves CP exactly, while a complex background of the $S$ field induces a transient bias among the degenerate CP vacua. Nevertheless, it remains unclear whether this mechanism can also remove topologically protected CP domain walls in the gauged-CP framework. We leave a detailed investigation of this case for a future study.

For the post-inflationary SCPV scenario, we naturally wonder the origin of baryon asymmetry of the Universe. At temperature higher than the CP-breaking scale $v_\text{CP}$, the VEV of $\eta$ vanishes, raising the question of what is the source of CP violation required for baryogenesis. In our setup, however, the $S$ field already carries a phase before the CP-breaking  transition occurs. While we have exploited this phase only as a source of a transient bias for CP domain wall decay, it is tempting to ask whether it could also participate in baryogenesis. For example, the phase of $S$ might provide a CP-violating source for thermal leptogenesis \cite{Fukugita:1986hr} even above the CP-breaking scale, while its time evolution could realize spontaneous baryogenesis via derivative couplings to baryon or lepton currents \cite{Cohen:1987vi,DeSimone:2016ofp}.
Furthermore, the collapse of CP domain walls occurs in an out-of-equilibrium environment in the presence of CP asymmetry, potentially opening a new scenario of domain-wall-mediated baryogenesis (see e.g. Refs.~\cite{Daido:2015gqa,Mariotti:2024eoh} for axion domain walls).
If realized, these possibilities would point toward a unified cosmological framework in which the transient CP-violating dynamics of $S$ are responsible for both the removal of CP domain walls, dark matter and the origin of baryon asymmetry.

\section*{Acknowledgments}

Y.N. is supported by Natural Science Foundation of Shanghai.
\bibliography{reference}

\end{document}